\documentclass{article}

\usepackage{PRIMEarxiv}

\usepackage[utf8]{inputenc} 
\usepackage[T1]{fontenc}    
\usepackage{hyperref}       
\usepackage{url}            
\usepackage{amsfonts}       
\usepackage{nicefrac}       
\usepackage{microtype}      
\usepackage{lipsum}
\usepackage{fancyhdr}       
\usepackage{graphicx}       

\usepackage{adjustbox}
\usepackage{calrsfs}
\usepackage{amssymb}
\usepackage{amsmath}
\usepackage{makecell} 
\usepackage{cite}
\usepackage{bm}
\usepackage{cases}
\usepackage{xcolor}
\usepackage{enumitem}
\usepackage{algpseudocode}
\usepackage{algcompatible}
\usepackage{algorithm}
\usepackage{caption}
\usepackage{booktabs}
\usepackage{multirow}
\usepackage{mathrsfs}
\usepackage{physics}
\usepackage{multicol}
\usepackage{float}
\usepackage{subcaption}
\floatstyle{ruled}
\newfloat{model}{thp}{lop}
\floatname{model}{Model}

\usepackage{multirow,mdframed}
\mdfsetup{skipabove=0pt,skipbelow=0pt}
\mdfdefinestyle{model}{%
  innertopmargin=0pt,
  innerbottommargin=0pt,
  innerleftmargin=0pt,
  innerrightmargin=0pt,
  skipbelow=0pt,%
  skipabove=0pt,
  splitbottomskip=0pt,
  splittopskip=0pt,
  leftmargin =0pt,%
    rightmargin=0pt,
  splittopskip=0pt,
    usetwoside=false,
}
\definecolor{coral}{RGB}{255,115,80}


\restylefloat{algorithm}

\graphicspath{{media/}}     

\begin{document}

\title{ Learning to Optimize meets Neural-ODE: Real-Time, Stability-Constrained AC OPF}

\author{
Vincenzo Di Vito\textsuperscript{*}\\
University of Virginia\\
\texttt{eda8pc@virginia.edu}
\And
Mostafa Mohammadian\textsuperscript{*}\\
University of Colorado Boulder \\
\texttt{mostafa.mohammadian@colorado.edu}
\And
Kyri Baker\\
University of Colorado Boulder \\
\texttt{kyri@colorado.edu}
\And
Ferdinando Fioretto \\
University of Virginia\\
\texttt{fioretto@virginia.edu} \\
}

\renewcommand{\thefootnote}{\fnsymbol{footnote}} 
\footnotetext[1]{Equal contribution.}

\renewcommand{\thefootnote}{\arabic{footnote}} 
\setcounter{footnote}{0} 


\maketitle

\begin{abstract}
Recent developments in applying machine learning to address Alternating Current Optimal Power Flow (AC OPF) problems have demonstrated significant potential in providing close to optimal solutions for generator dispatch in near real-time.
While these \emph{learning to optimize} methods have demonstrated remarkable performance on steady-state operations, practical applications often demand compliance with dynamic constraints when used for fast-timescale optimization.
This paper addresses this gap and develops a \emph{real-time stability-constrained OPF} model (DynOPF-Net) that simultaneously addresses both \emph{optimality} and \emph{dynamical stability} within learning-assisted grid operations. The model is a unique integration of learning to optimize that learns a mapping from load conditions to OPF solutions, capturing the OPF's physical and engineering constraints, with Neural Ordinary Differential Equations, capturing generator dynamics, enabling the inclusion of a subset of stability constraints.
Numerical results on the WSCC 9-bus and IEEE 57-bus benchmark systems demonstrate that DynOPF-Net can produce highly accurate AC-OPF solutions while also ensuring system stability, contrasting the unstable results obtained by state-of-the-art LtO methods.
\end{abstract}



\section{Introduction}

The AC Optimal Power Flow (AC-OPF) problem plays a fundamental role in power systems as it determines the optimal generator dispatch to meet demands while adhering to physical and engineering constraints. Traditionally, OPF problem instances are solved on timescales spanning 5 minutes to 1 hour, and often utilize linear approximations which do not satisfy steady-state AC power flow~\cite{bakerOPF21}. Additionally, the stochastic nature of renewable energy sources has increased uncertainty, necessitating more accurate power flow representations, and more frequent adjustments by system operators to preserve system stability~\cite{8756047}. Currently, these operations typically rely on heuristics for maintaining stable frequencies and voltages of the electrical generators, which lead to inefficiencies and higher losses. Rapid fluctuations of intermittent renewable generation, for example, can lead to curtailments as well as increasing emissions~\cite{Mlilo2021}. Similarly, governor droop control, which uses proportional control rules to modify the power generator based on frequency imbalances, introduces additional inefficiencies, losses, and emissions~\cite{6702462}. 

There are two key challenges within this setting. The {\em first} is the complexity of solving the full AC-OPF problem, which limits the frequency of these operations, 
and the {\em second} is the dynamic nature of the system, which impacts global system stability~\cite{9286772}. Indeed, relying solely on steady-state models for power dispatching operations is insufficient to guarantee compliance with system dynamic requirements. Nonetheless, adhering to these dynamics is not only computationally demanding on its own, but fully incorporating them within the AC-OPF model would introduce complex interdependencies among the problem variables, significantly increasing the computational complexity of an already challenging problem. 

In an effort to address the first challenge, there has been recent interest in applying machine learning (ML) based models to learn solutions mappings from load conditions to AC-OPF solutions~\cite{9940481}.
These approaches have demonstrated enormous potential for replacing traditional numerical solvers to approximate complex supply/demand balance problems in near real-time. 
Central to these advancements is the concept of \emph{\LtO} (LtO),~\cite{kotary2021end}
where neural network-based models serve as proxies for traditional numerical solvers to produce near-optimal solutions for parametric constrained optimization problems. 
However, while LtO approaches have shown promising results in solving the AC-OPF problem with high fidelity in real-time~\cite{fioretto2020predicting, donti2020dc3}
, they address the \emph{steady-state} problem, and may not be suitable to capture the dynamics of the system.
In Section \ref{sec: experimental result} the paper will show that, on the benchmark systems analyzed, many of the solutions generated by state-of-the-art LtO methods for AC-OPF violate the dynamic requirements of a synchronous electrical generator, which is unacceptable in practice.


To overcome these challenges, this paper proposes \emph{Dynamic OPF-Net} (DynOPF-Net), a learning-based model that combines \LtO{} with ML-based dynamic predictors to solve real-time AC OPF problems while also addressing the dynamic stability of the generators.
A key component of DynOPF-Net is the integration of neural ODEs (NODEs)~\cite{kidger2022neural} within the \LtO{} framework. NODEs approximate the continuous dynamics of systems through neural networks, allowing efficient modeling of system behaviors that evolve over time. 
The paper shows how this integration is able to produce near-optimal and stable 
solutions 
for a variety of operational conditions.

\textbf{Contributions.}
This paper offers the following key contributions:
{\bf (1)} We introduce Dynamic OPF-Net (DynOPF-Net), a novel framework that seamlessly integrates machine learning with optimization techniques to incorporate generator dynamics directly into the AC-OPF model. 
{\bf (2)} We provide empirical evidences which highlight the critical need to model the system dynamics within the OPF framework. Specifically, we demonstrate that existing LtO methods, which overlook these dynamics, systematically fail to satisfy stability requirements. In contrast, DynOPF-Net consistently generates solutions that adhere to these dynamic constraints, helping address stability. 
{\bf (3)} We show that DynOPF-Net not only achieves a decision quality comparable to that of state-of-the-art LtO methods that address only steady-state constraints but also uniquely ensures compliance with dynamic requirements. This is achieved without sacrificing computational efficiency, offering the first solution, to our knowledge, for real-time power system operations under dynamic conditions.


\section{Related Work}\label{sec:r elated work}

\paragraph{Learning methods for OPF}
In recent years, numerous ML-based approaches have been proposed to replace traditional OPF numerical solvers. Among the first attempts,~\cite{ng2018statistical} uses a statistical learning-based approach to predict DC-OPF solutions, while~\cite{9612004} proposes a DNN informed by a sensitivity to learn OPF solutions, which requires computing the sensitivities of OPF solvers with respect to load variables. Despite the promising results, these approaches did not focus on constraint satisfaction and thus may produce many infeasible solutions. 

These deficiencies lead to developing methods that integrate the OPF problem structure within deep learning-based models, giving rise to \emph{physics informed} or \emph{learning to optimize} methods for OPF\footnote{Here, we are going to use the latter term, as physics informed neural networks refers to another concept used to approximate PDEs with neural networks, as we will discuss later in this paper.}.
In particular,~\cite{Deka2019LearningFD} uses a DNN to identify the active constraint sets to simplify and enhance learning DC OPF solution mappings.  
A Lagrangian-Dual deep learning-based approach was introduced in~\cite{fioretto2020predicting}, which aims to learn near-optimal solutions while also encouraging satisfaction of the OPF constraints. Their approach modifies the neural network training by parameterizing the loss function with constraint penalty terms proportional to the degree of the predicted constraint violations, mimicking a dual ascent method~\cite{boyd2011distributed}. Similarly, the method proposed in this paper uses dual penalty terms to drive the learning model towards solutions which are feasible and stable. 
There has been since then a number of approaches developing on and improving these techniques, including recurrent architectures~\cite{MostafaRNN} and decomposition methods~\cite{chatzos2021spatialnetworkdecompositionfast}. 
While these methods produce state-of-the-art results for AC OPF, 
they do not guarantee strict compliance with feasibility requirements.
Hence, recent work in ML-driven OPF solvers focuses on ensuring constraint satisfaction. Specifically,~\cite{Zamzam_learn_19} predicts partial OPF solutions and subsequently resolve the remaining variables by addressing balance flow equations. In~\cite{donti2020dc3}, this approach is built upon by employing implicit layers, enabling the correction of inequalities and the completion of equality constraints within the training process. 

While these works have shown that machine learning methods can produce near-optimal OPF solutions, they focus on the \emph{steady-state} dispatch problem, ignoring power system dynamics. This fundamentally limits applicable timescales (and speed improvements) from these methods. To the best of our knowledge, this paper is the first to introduce a machine learning-based method for solving AC OPF while simultaneously integrating synchronous generator stability.

\paragraph{Learning methods for system dynamics}
The system dynamics typically take the form of a set of Ordinary (ODEs) or Partial Differential Equations (PDEs), which describe the laws governing the state variables of the system. When the number of ODE variables is large, the computational complexity renders precise numerical solutions impractical for real-time applications~\cite{SU20212113}, 
where instead highly accurate approximations of the system of ODEs (PDEs) are preferred. 

In this context, several works proposed physics-informed neural networks (PINNs)~\cite{2019JCoPh.378..686R} to embed governing equations within the training of a ML model. PINNs have been shown highly effective in approximating various systems of complex PDEs at extremely fast inference times.
In the power system literature, PINNs have mainly been used in single-machine infinite bus systems to estimate power system state variables and unknown parameters~\cite{Misyris2019PhysicsInformedNN}. 
While PINNs offer significant modeling advantages by integrating physical behaviors within the model, they face training stability challenges when applied to complex systems, such as power grids~\cite{NEURIPS2021_df438e52}. 
In addition, as PINNs are designed to approximate the solution of \emph{specified} system of differential equations, they suffer from limited generalization capability~\cite{JMLR:v24:21-1524}.  
This severely limits their applicability in the context studied in this paper where the system dynamics can vary, and thus neural surrogates are required to capture \emph{family} of dynamics. 

To address this challenge, researchers have recently developed neural {differential equations}~\cite{kidger2022neural}. Among those, neural ODEs 
learn the system dynamics by approximating 
{the vector field with a neural network}. This capability makes NODEs well-suited as dynamic predictors for parametric systems of ODEs. In the context of power system dynamics, this allows for stability analysis under various operational conditions.
In the power system literature, NODEs have been adopted for inferring critical state information of the power system dynamics~\cite{9844253}. 
To the best of the authors' knowledge, ours is the first attempt to integrate NODE models as dynamic predictors within the AC OPF problem.

\section{Problem Formulation}\label{sec: prob_formulation}

This section introduces the stability-constrained AC OPF problem, starting from two key components: the (steady state) AC OPF problem and the synchronous generator dynamics.  The paper adopts the following notation: lowercase symbols are used for scalars, bold symbols represent vectors, 
and uppercase symbols 
represent complex variables, denoted either in rectangular or polar form. 
Sets are denoted with standard calligraphic symbols (e.g., \(\pazocal{X} = \{x_1, \dots, x_n\}\)), and special calligraphic symbols are reserved for models, such as a deep neural network parameterized by vector \(\bm{\phi}\), denoted as \(\mathcal{X}_{\bm \phi}\).

\smallskip\noindent{\textbf{AC Optimal Power Flow.}}
The AC OPF problem determines the most cost-effective generator dispatch that satisfies demand within a power network subject to various physical constraints. Typically, the setting focuses on a snapshot of the OPF problem in time. 
In a power network, represented as a graph $(\pazocal{N,L})$, the node set $\pazocal{N}$ consists of \(n\) buses, and the edge set \(\pazocal{L}\) comprises \(l\) lines. 
Additionally, $\pazocal{G}$ is used to represent the set of synchronous generators in the system. 
The AC power flow equations use complex numbers for current \(I\), voltage \(V\), admittance \(Y\), and power \(S\), interconnected through various constraints. 
The power generation and demand at a bus \(i \in \pazocal{N}\) are represented by \(S^r_i = p_i^r + jq_i^r\) and \(S^d_i = p_i^d + j q_i^d\), respectively, while the power flow across line \(ij\) is denoted by \(S_{ij}\), and \(\theta_i\) describes the phase angles at bus \(i \in \pazocal{N}\).
The Kirchhoff's Current Law (KCL) is represented by \( I^r_i - {I^d_i} = \sum_{(i,j)\in \pazocal{L}} I_{ij}\), Ohm's Law by \( I_{ij} = {Y}_{ij} (V_i - V_j)\), and AC power flow denoted \( S_{ij} = V_{i}I_{ij}^\ast\). 
These principles form the AC Power Flow equations, described by ~\eqref{eq:ac_3} and \eqref{eq:ac_4} in Model~\ref{model:ac_opf}. The goal is to minimize a function~\eqref{ac_obj} representing dispatch costs for each generator.
Constraints \eqref{eq:ac_1}-\eqref{eq:ac_6} represents voltage operational limits to bound voltage magnitudes and phase angle differences, while \eqref{eq:ac_2}-\eqref{eq:ac_5} set boundaries for generator output and line flow. Constraint~\eqref{eq:ac_7} sets the reference phase angle. Finally, constraints~\eqref{eq:ac_3} and \eqref{eq:ac_4} enforce KCL and Ohm's Law, respectively. 

Note that formulation alone does \emph{not} capture synchronous generator dynamics, although stability-aware linearized OPF formulations have been developed~\cite{Hafez16}. 
\begin{model}[!t]
    \caption{The AC Optimal Power Flow Problem (AC-OPF)}
    \label{model:ac_opf}
    \vspace{-9pt}
    {\small
    \begin{subequations}
    \begin{align}
        \mbox{\bf variables:} \;\;
        & S^r_i, V_i \;\; \forall i\in \pazocal{N}, \;\;
          S_{ij}     \;\; \forall(i,j)\in \pazocal{L} \nonumber \\
        \mbox{\bf minimize:} \;\;
        & \sum_{i \in \pazocal{G}}  c_{2i} (\Re(S^r_i))^2 + c_{1i}\Re(S^r_i) +  c_{0i} \label{ac_obj} \\
        \mbox{\bf subject to:} \;\; 
        & v^l_i \leq |V_i| \leq v^u_i       \;\; \forall i \in N \label{eq:ac_1} \\
        & -\theta^\Delta_{ij} \leq \angle (V_i V^*_j) \leq \theta^\Delta_{ij} \;\; \forall (i,j) \in \pazocal{L}  \label{eq:ac_6}  \\
        & S^{rl}_i \leq S^r_i \leq S^{ru}_i \;\; \forall i \in \pazocal{N} \label{eq:ac_2}  \\
        & |S_{ij}| \leq s^u_{ij}                  \;\; \forall (i,j) \in \pazocal{L} \label{eq:ac_5}  \\
        & S^r_i - S^d_i = \textstyle\sum_{(i,j)\in L} S_{ij} \;\; \forall i\in \pazocal{N} \label{eq:ac_3}  \\ 
        & S_{ij} =  Y^*_{ij} |V_i|^2 - Y^*_{ij} V_i V^*_j             \;\; \forall (i,j)\in \pazocal{L}
        \label{eq:ac_4} \\
        &\theta_{\text{ref}} = 0 \label{eq:ac_7}
    \end{align}
    \end{subequations}
    }
    \vspace{-14pt}
\end{model}

\smallskip\noindent{\textbf{Synchronous Generator Model.}}
To capture the dynamic behavior of synchronous generators with high fidelity, 
this study considers the \emph{classical} ``machine model''~\cite{sauer1998power}, a simplified version of the ``two-axis model''. The dynamics of a synchronous generator $g \in \pazocal{G}$ are defined as:
\begin{align}
\label{eq:2}
    &\frac{d}{dt}\!\begin{bmatrix}
    e^{\prime g}_q (t)\\
    e^{\prime g}_d (t)\\
    \delta^g (t) \\
    \omega^g (t)
    \end{bmatrix}
    \!\!
\!=\!
\begin{bmatrix}
0 \\
0 \\
\omega_s (\omega^g(t) - \omega_s) \\
\frac{1}{m^g} (p_m^g - e^{\prime g}_d(t) i_d^g(t) - e^{\prime g}_q(t) i_q^r(t) - d^g(\omega^g(t) - \omega_s))
\end{bmatrix},
\end{align}
where stator currents \( i_{d}^g(t) \) and \( i_{q}^g(t) \) in the reference frame of machine $g \in \pazocal{G}$, are computed as:
\begin{align}
\label{eq: stator_current}
    \begin{bmatrix}
i_d^g(t) \\
i_q^r(t)
\end{bmatrix}
\!\! \! =\!
\begin{bmatrix}
0 \!\!\!&\!\!\! -x^{\prime g}_d \\
x^{\prime g}_d \!\!\!&\!\!\! 0
\end{bmatrix}^{-1}
\begin{bmatrix}
e^{\prime g}_d(t) - |V_g| \sin(\delta^g(t) - \theta_g) \\
e^{\prime g}_q(t) - |V_g| \cos(\delta^g(t) - \theta_g)
\end{bmatrix}.
\end{align}

\noindent
For a generator $g\in \pazocal{G}$, the state vector  $\begin{bmatrix} e^{\prime g}_{q} & e^{\prime g}_{d} & \delta^g & \omega^g \end{bmatrix}^\top$ describes its voltage components, \( e^{\prime g}_q \) and \( e^{\prime g}_d \), associated with the `d' and `q' axis (\textit{dq}-axes of a reference frame for $g$), its rotor angle \( \delta^g \), and its angular frequency \(\omega^g\). The synchronous angular frequency is denoted as $\omega_s$ in Equation \eqref{eq:2}.
The mechanical power \( p_{m}^g \) from the shaft turbine serves as the control input. 
Constitutive machine parameters are given by \(\begin{bmatrix} x^{\prime g}_{d} & m^g & d^g \end{bmatrix}^T
\), where \( x^{\prime g}_{d} \) denotes the the transient reactance, \( m^g \) the machine's inertia constant and \( d^g \) the damping coefficient.
Importantly, stator currents are connected to the terminal voltage magnitude \( |V_g| \) and phase angle \( \theta_g \) via~\eqref{eq: stator_current}.
In this model, voltage components \( e^{\prime g}_d(t) \) and \( e^{\prime g}_q(t) \) are held constant, meaning the update function for these states is zero.
The interested reader is referred to~\cite{sauer1998power} for a more in-depth discussion of the classical machine model.
Given the assumptions on the voltage \( e^{\prime g}_d \) and \( e^{\prime g}_q \), the dynamic model of synchronous generators results in:
\begin{equation}
\mkern-18mu \frac{d}{dt}
\begin{bmatrix}
\delta^g (t)\\
\omega^g (t)
\end{bmatrix}
\!\!
\!=\!
\begin{bmatrix}
\omega_s (\omega^g(t) - \omega_s) \\
  \frac{1}{m^g} \!\left( p_m^g - d^g(\omega^g\!(t) - \omega_s) - \frac{e^{\prime g}_{q}(0) |V_g|}{x^{\prime g}_d} \sin(\delta^g\!(t) - \theta_g) \right) 
\end{bmatrix}\!.\!
\label{eq: generator dynamics}
\end{equation}

\smallskip\noindent{\textbf{Initial Values of Rotor Angles and EMF Magnitudes.}}
For each generator $g \in \pazocal{G}$, the initial values of rotor angle $\delta^g(0)$ 
and electromotive force (EMF) $e^{\prime g}_{q}(0)$ are derived from the active and reactive power equations, assuming the dynamical system~\eqref{eq: generator dynamics} initially being in a steady state condition, $\frac{d}{dt}\begin{bmatrix}\delta^g(t) & \omega^g(t) \end{bmatrix}_{t=0}^T = \begin{bmatrix}0 & 0\end{bmatrix}^T$, from which: 
\begin{align}
    &\frac{e^{\prime g}_{q}(0) |V_g| \sin(\delta^g(0) - \theta_g)}{x^{\prime g}_d}  - p^r_g = 0,  \label{eq: init_delta}\\
    & \frac{e^{\prime g}_{q}(0) |V_g| \cos(\delta^g(0) - \theta_g) - |V_g|^2}{x^{\prime g}_d} - q^r_g = 0. \label{eq: init_EMF}
\end{align}
Following the same assumption, it follows that
\begin{align}
   \omega^g(0) = \omega_s. \label{eq: init_omega}
\end{align}

\smallskip\noindent{\bf Stability Limit.}
To guarantee stability of a synchronous generator $g \in \pazocal{G}$, the rotor angle $\delta^g(t)$ is required to remain below an instability threshold $\delta^{\max}$, as defined by SIngle Machine Equivalent (SIME):
\begin{align} \label{eq: stab-limit}
    \delta^g(t) \leq \delta^{\max} \quad \forall t\geq 0.
\end{align}
Unstable conditions arise when violating~\eqref{eq: stab-limit}, which is the principal binding constraint that necessitates re-dispatching.

\begin{figure*}[t!]
\centering
\includegraphics[width=0.9\linewidth]{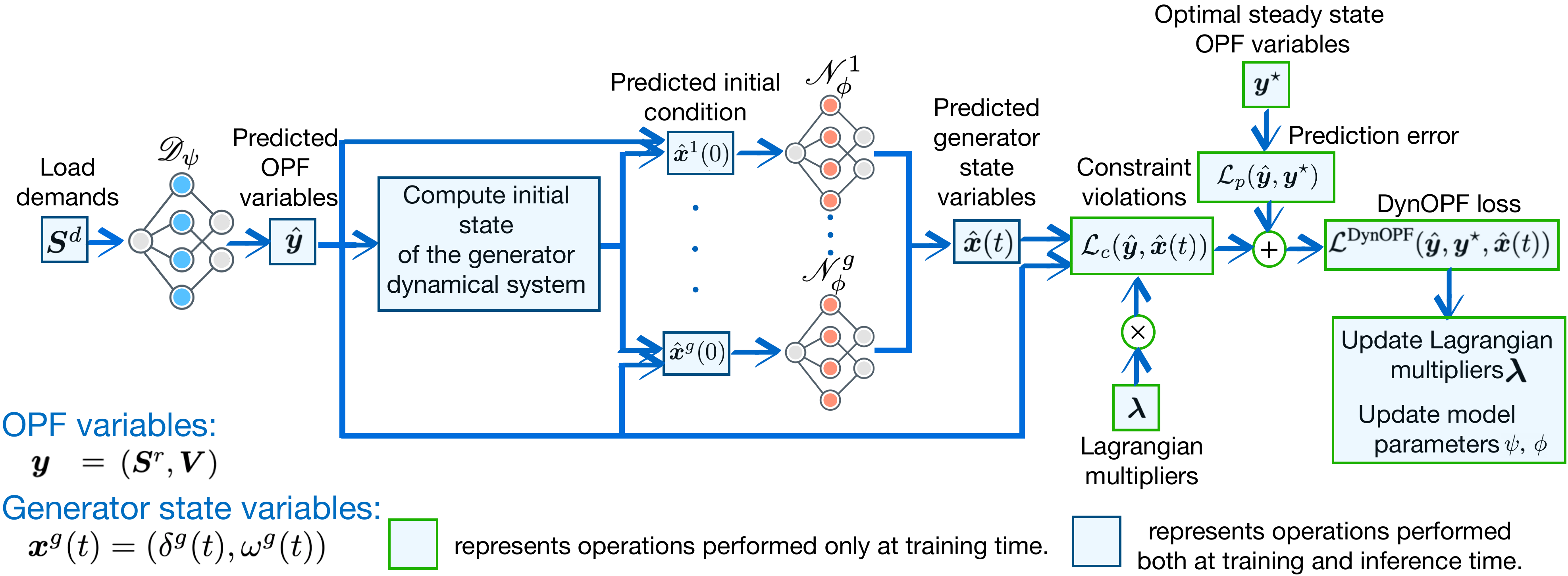}
\caption{DynOPF-Net uses a dual network architecture consisting of a LtO model $\mathcal{D}_\psi$ which takes as input a load demand 
$\bm{S}^d$ 
and output an estimate the OPF variables $\hat{\bm y}$, based on which neural-DE models $\mathcal{N}^g_\theta$ output the generators state-variables $\hat{\bm x}^g(t)$. These estimates are used to compute the constraint violations $\pazocal{L}_{c}(\hat{\bm y}, \hat{\bm{x}}(t))$ and the prediction error $\pazocal{L}_p(\hat{\bm y}, \bm y^\star)$ terms, which sum constitutes the total loss function $\pazocal{L}^{\text{DynOPF}}(\hat{\bm y}, \bm y^\star, \hat{\bm{x}}(t))$.}
\label{fig: DynOPF-Net}
\end{figure*}

\smallskip\noindent{\textbf{Stability-Constrained OPF.}}
\label{ssec:stability_constrained_OPF}
The \emph{stability-constrained} OPF problem is thus formulated in Model~\ref{model:stability_constrained_ac_opf}. This problem determines the optimal power dispatch subject to physical, engineering, and dynamic stability constraints~\eqref{eq:scac_1_7}-\eqref{eq:scac_13}. Given a load demand 
$\bm{S}^d$
, the objective is to find the OPF variables 
$\bm S^r, \bm V$
that minimize the cost function~\eqref{scac_obj} while satisfying the constraints~\eqref{eq:scac_1_7}-\eqref{eq:scac_13}. The inclusion of the dynamic behavior of synchronous generators complicates the problem due to the coupling between the OPF variables $S_g^r, V_g$ 
and the generator state variables $\delta^g(t),  \omega^g(t)$, as shown in~\eqref{eq:scac_9}-\eqref{eq:scac_11}. Additionally, the non-linear nature of the generator model and the time dependencies of the state variables further increase the complexity. As a result, traditional numerical optimization algorithms are unable to handle, in a computationally viable way, the stability-constrained AC-OPF. To address this challenge, we develop a learning-based dual framework integrating NODEs with LtO models.

\begin{model}[!t]
    \caption{The Stability Constrained AC-OPF Problem}
    \label{model:stability_constrained_ac_opf}
    \vspace{-9pt}
    {\small
    \begin{subequations}
    \begin{align}
        \mbox{\bf variables:} \;\;
        & S^r_i, V_i \;\; \forall i\in \pazocal{N}, \;\; \delta^g(t), \omega^g(t) \;\; \forall g\in \pazocal{G}, \;\; \nonumber
         \\ & S_{ij}     \;\; \forall(i,j)\in \pazocal{L} \nonumber \\
        \mbox{\bf minimize:} \;\;
         & \sum_{i \in \pazocal{G}}  c_{2i} (\Re(S^r_i))^2 + c_{1i}\Re(S^r_i) +  c_{0i} \label{scac_obj} \\
        \textbf{subject to:} \;\;
        & (\ref{eq:ac_1}) \text{ -- } (\ref{eq:ac_7}) \label{eq:scac_1_7}\\
        &\hspace{-20pt}
        \frac{d \delta^g(t)}{dt}  = \omega_s (\omega^g(t) - \omega_s) \;\; \forall g \in \pazocal{G}
        \label{eq:scac_8}\\
        &\hspace{-20pt}
        \frac{d \omega^g(t)}{dt} = \frac{1}{m^g} \left( p_m^g - d^g(\omega^g(t) - \omega_s) \right) \nonumber\\
        & - \frac{e^{\prime g}_{q}(0) |V_g|}{x^{\prime g}_d m^g} \sin(\delta^g(t) - \theta_g) \;\; \forall g \in \pazocal{G}
        \label{eq:scac_9}\\
        &\hspace{-20pt}
        \frac{e^{\prime g}_{q}(0) |V_g| \sin(\delta^g(0) - \theta_g)}{x^{\prime g}_d }  - p^r_g = 0 \;\; \forall g \in \pazocal{G}  \label{eq:scac_10}\\
        &\hspace{-20pt}
        \frac{e^{\prime g}_{q}(0) |V_g| \cos(\delta^g(0) - \theta_g) - |V_g|^2}{x^{\prime g}_d} - q^r_g = 0 \;\; \forall g \in \pazocal{G} \label{eq:scac_11}\\
        &\hspace{-20pt}
        \omega^g(0) = \omega_s \;\; \forall g \in \pazocal{G} \label{eq:scac_12}\\
        &\hspace{-20pt}
        \delta^g(t) \leq \delta^{\max} \;\; \forall g \in \pazocal{G} . \label{eq:scac_13}
    \end{align}
    \end{subequations}
    }
    \vspace{-14pt}
\end{model}

\section{Dynamic OPF-Net} 
\label{sec: methods}

Given a load profile $\bm{S}^d = (\bm p^d, \bm q^d)$, the goal is to \emph{predict} the generators set-points 
that minimize the objective~\eqref{scac_obj} while simultaneously satisfying the operational and dynamic constraints~\eqref{eq:scac_1_7}-\eqref{eq:scac_13} of Problem~\ref{model:stability_constrained_ac_opf}. 
A significant challenge in this learning task is ensuring the satisfaction of both steady-state~\eqref{eq:scac_1_7} and dynamic constraints~\eqref{eq:scac_13}. 
To address this challenge, we propose \emph{Dynamic OPF-Net} (DynOPF-Net), a dual neural network architecture consisting of a \textcolor{cyan}{learning-to-optimize model} (cyan network in Fig.~\ref{fig: DynOPF-Net}) and a collection of \textcolor{coral}{NODE models}, one for each generator, as illustrated in the orange\textbf{} networks of Fig.~\ref{fig: DynOPF-Net}.
The LtO model, denoted as $\mathcal{D}_{\psi}$, where $\psi$ represents its parameters, predicts the OPF variables $\bm{S}^r = (\bm p^r, \bm q^r)$ and $\bm V$.
Each NODE model $\mathcal{N}^g_{\phi}$, parametrized by $\phi$, captures the system dynamics of the corresponding generator $g \in \pazocal{G}$. Both the steady-state and dynamic constraints are integrated into the learning task using a Lagrangian Dual Learning framework~\cite{fioretto2020predicting} as reviewed in Section \ref{sec:DynOPF-Net}. 
\subsection{Approximating Synchronous Generators Dynamics}
\label{sec: NODE}

This section first focuses on approximating the dynamics of synchronous generators (see Equation~\eqref{eq: generator dynamics}) using NODE models~\cite{kidger2022neural}. This proxy for a numerical differential solver allows us to infer generator state variables $\delta^g(t), \omega^g(t)$ for each generator $g$ at fast computational times. 

Consider a generic ODE,
\begin{equation}
\centering
\begin{cases}
\label{eq:generic_ode}
\frac{d\bm{x}(t)}{dt} = \bm{p}(\bm{x},t) \\
\bm{x}(0) = \bm{x}_0,
\end{cases}
\end{equation}
where $\bm{x} \in \mathbb{R}^n$ describes the $n$-dimensional vector of state variables, $\bm{x}_0 \in \mathbb{R}^n$ the initial conditions, $t \geq 0$ the time, and $\bm{p} \colon \mathbb{R}^n \times \mathbb{R}_+ \to \mathbb{R}^n$ a vector field. A neural ODE defines the continuous dynamics of system~\eqref{eq:generic_ode} by
replacing the vector field $\bm{p}(\bm{x},t)$ with a neural network $\mathcal{N}_{\phi}(\bm{x},t)$, such that 
\(
\bm{p}(\bm{x},t) \approx \mathcal{N}_{\phi}(\bm{x},t).
\)

From a practical standpoint, the forward pass of a NODE model is computed using a numerical algorithm similar to how a traditional ODE solver computes the numerical solution (by iteratively applying an update step starting from the initial condition $\bm{x}(0)$). However, instead of evaluating the true vector field  $\bm{p}$, it uses the neural network  $\mathcal{N}_{\phi}$. 
This approximation simplifies the computation of the update step in a numerical ODE solver, as it does not require evaluating the vector field $\bm p$ but only its approximation $\mathcal{N}_\phi$, thus enabling fast computation of the forward pass \cite{kidger2022neural}. 
This results in a significant speed advantage over traditional ODE solvers that use the true $\bm{p}$.

Training a NODE $\mathcal{N}_\phi$, involves a dataset $\pazocal{D}$ of pairs $(\bm{x}_0, \bm{x}(t))$ to optimize the following empirical risk problem:
\begin{subequations}
    \label{eq:NODE_training}
    \begin{align}
    \label{eq:NODE_loss}
        \minimize_{\phi} & \; \; \mathbb{E}_{(\bm{x}_0, \bm{x}(t)) \sim \pazocal{D}} ||\hat{\bm{x}}(t) - \bm{x}(t)||^2 \\
        s.t. & \; \;  \hat{\bm{x}}(t) = \textsl{ODEsolver}(\mathcal{N}_{\phi}, \bm{x}_0, \Delta_t ) \label{eq:NODE_forward}\\
             & \; \;  {\bm{x}}(t) = \textsl{ODEsolver}(\bm p, \bm{x}_0, \Delta_t ).\label{eq:ODE_forward}
    \end{align}
\end{subequations}
Therein, \textsl{ODEsolver} denotes a numerical algorithm such as Euler's implicit and $\Delta_t$ is the integration time step.
In this paper, NODE models are used to learn the solutions of a family of generator dynamics given by Equation~\eqref{eq: generator dynamics}. They can be trained as described in~\eqref{eq:NODE_training}, with the difference that the numerical solver solutions  $\bm{x}(t)$  are obtained from different instances of the generator model by varying the OPF variables which influence the initial conditions $\bm{x}_0$ and governing equations~\eqref{eq: generator dynamics} (denoted with $\bm{p}$ in \eqref{eq:ODE_forward}). 
Due to the interaction between the generator state variables $\delta^g(t), \omega^g(t)$ and the OPF variables 
$|V_g|, \theta_g$,
it is useful to express the NODE estimates as a function of the initial conditions $\bm{x}^g(0)=(\delta^g(0), \omega^g(0))$ and the estimated OPF voltages variables $\hat{V}_g, \hat{\theta}_g$ of Model \ref{model:stability_constrained_ac_opf}:
\begin{equation}
\label{eq: NODE dynamics predictors}
   \hat \delta^g(t), \hat \omega^g(t) 
   = \mathcal{N}_{\phi}^g(\hat \delta^g(0), \hat \omega^g(0), |\hat V_g|, \hat \theta_g).
\end{equation}
Here $|\hat V_g|$ and $\hat \theta_g$ are the (approximate) voltage magnitude and angles associated with bus of generator $g$ as predicted by a learning-to-optimize model $\mathcal{D}_{{\psi}}(\bm{S}^d)$, as discussed in the following section. Note that, as these predicted values depend on the particular stability-constrained OPF problem instance and are not known a priori, the NODE model has to be able to produce accurate state variable estimates for a \emph{range} of predicted $|\hat V_g|$, $\hat \theta_g$ values, which (feasible) bounds are given by Constraints~\eqref{eq:ac_1} and~\eqref{eq:ac_6}. Given these predicted values, the estimated initial state variables $\hat \delta^g(0)$ and $\hat \omega^g(0)$ are instead defined by Equations~\eqref{eq: init_delta}--\eqref{eq: init_omega}. 
Note that the estimated OPF quantities $|\hat V_g|$ and $\hat \theta_g$ not only influence the initial state variables $\delta^g(0), \omega^g(0)$, but also impact the governing equation of the generator model, formalized in Equation~\eqref{eq: generator dynamics}.

Equation~\eqref{eq: NODE dynamics predictors} implies that we consider an \emph{augmented} generator model, where two additional state variables, $|V_g|(t)$ and $\theta_g(t)$, with no dynamics (i.e., $\frac{d |V_g|(t)}{dt}=0, \frac{d \theta_g(t)}{dt}=0$), and initial condition $|V_g|(0)=|\hat{V}_g|, \theta_g(0)=\hat{\theta}_g$ are incorporated alongside the actual state variables $\delta^g(t)$ and $\omega^g(t)$. This approach allows us to explicitly inform the NODE of the role played by $V_g$ on the dynamics of each generator.

This setup establishes the foundation for training $\mathcal{N}_{\phi}^g$ to learn a family of generator dynamics, defined by different 
\emph{extended} initial condition vector $[  
\hat \delta^g(0) \ \hat \omega^g(0) \ |\hat V_g| \ \hat \theta_g]^T$ and different instances of its governing equations.

\subsection{Learning Stability-constrained ACOPF Solutions}
\label{sec:DynOPF-Net}
We are now ready to discuss the interaction between the Neural ODEs $\mathcal{N}_\phi^g$ capturing the generators 
dynamics and the learning to optimize model $\mathcal{D}_\phi$ that predicts the OPF dispatch values. 
The synergistic integration of these components allows us to incorporate both static and dynamic constraints into the learning process, ensuring that the predicted solutions satisfy the stability-constrained AC OPF 
requirements. 


\noindent Training this learning task involves a dataset $\{\left( \bm{S}^{d,i}, \bm{y}^{\star,i} \right)\}_{i=1}^{n_\text{obs}}$, where each of the $n_\text{obs}$ data points describes the observations of load demands ($\bm{S}^{d}$) and the corresponding optimal values of the OPF variables $\bm{y}^{\star} = \left( \bm{S}^{\star,r}, {\bm V}^{\star} \right)$, under the assumption that all synchronous generators are at a \emph{steady-state}.

As shown in Figure~\ref{fig: DynOPF-Net}, given a load demand $\bm{S}^d$, the DynOPF-Net' LtO model $\mathcal{D}_{{\psi}}$ produces an estimate of the OPF variables $\hat{\bm y} = \mathcal{D}_{{\psi}}(\bm{S}^d)$. 
To ease notation, in the following, we denote with $\hat{\bm x}^g(t) = \mathcal{N}^g_{\phi}(\hat{\bm{x}}^g(0), \hat{V}_g)$ the NODE estimate of the generator $g$'s state variables, highlighting they are a function of the estimated initial conditions $\hat{\bm{x}}^g(0)$ and predicted voltage $\hat{V}_g)$ (see Equation~\eqref{eq: NODE dynamics predictors}). 
Additionally, $\bm{x}(t)$, denotes all the generators' state variables, while $\bm{\hat{x}}(t)$ denotes the state variable estimated by NODE models $\bm{\mathcal{N}}_{\phi}^g$ associated with each generator $g$ in the network.


The dynamics learned by each NODE are integrated with the LtO predictions exploiting a Lagrangian dual framework \cite{fioretto2020predicting}. For a given constrained optimization problem, in Lagrangian duality, a subset of the problem’s constraints is relaxed into the objective function, each associated with a multiplicative weight called a \emph{Lagrange multiplier}. This modified objective function is known as the \emph{Lagrangian dual} function. To find the best Lagrange multipliers, the framework solves a max-min problem that seeks the optimal multipliers while minimizing the associated Lagrangian dual function. In the proposed deep learning framework, we use a similar technique to integrate the ``stability constraints'' into the learning process. 
Within this framework, the LtO model may also employ a Lagrangian approach to incorporate the OPF constraints~\eqref{eq:ac_1}--\eqref{eq:ac_7} directly into the optimization process, depending on the specific learning scheme adopted (the various LtO schemes adopted are 
 elaborated in the Experimental setting, Section \ref{sec: experimental setting}).

The augmented Lagrangian loss function incorporates both the prediction error $\pazocal{L}_p$ and penalty terms for constraint violations $\pazocal{L}_c$:
\begin{equation}
\label{eq:LD_loss}
\pazocal{L}^{\text{DynOPF}}(\hat{\bm y}, \bm y^\star, \hat{\bm{x}}(t)) =  \pazocal{L}_p(\hat{\bm y}, \bm y^\star) + \pazocal{L}_{c}(\hat{\bm y}, \hat{\bm{x}}(t)),
\end{equation}
where
\begin{equation}
\label{eq:LD_loss_pred}
    \pazocal{L}_p(\hat{\bm y}, \bm y^\star) = \|\hat{\bm y}-\bm y ^\star\|^2,
\end{equation}
represents the prediction error (MSE), with respect to the \emph{steady-state} optimal OPF variable $\bm y ^\star$ and
\begin{equation}
\label{eq:LD_loss_constr}
\begin{split}
    \pazocal{L}_c(\hat{\bm y}, \hat{\bm{x}}(t)) = &\sum_{j=1}^{n_{\text{eq}}} {\lambda}_{h_j} \nu \left( h_j(\hat{\bm y}, \hat{\bm{x}}(t)) \right)+ \sum_{l=1}^{n_{\text{ineq}}} {\lambda}_{u_l} \nu \left(u_l(\hat{\bm y}, \hat{\bm{x}}(t)) \right)
\end{split}
\end{equation}
is a weighted sum of constraint violations incurred by the DynOPF-Net predictions $\hat{\bm y}$, $\hat{\bm x}(t)$. Here $\nu$ is a differentiable function which computes the amount of violation of a given constraint (e.g. for a linear inequality constraint $ay \leq b$, the corresponding violation returned by $\nu$ is given by $\max(0,ay-b))$, while for a linear equality constraint $ay = b$, the violation returned by $\nu$ is $|ay-b|$).
Functions $h_j, \; j=1, \ldots, n_{eq}$ denote the static equality constraints in \eqref{eq:scac_1_7} and dynamic constraints \eqref{eq:scac_9}-\eqref{eq:scac_12} of Model \ref{model:stability_constrained_ac_opf}, where 
constraints \eqref{eq:scac_9}, \eqref{eq:scac_10} and \eqref{eq:scac_12} are written in an implicit form.
Functions $u_l, \; l=1,\ldots, n_{ineq}$ denote the static inequality constraints in \eqref{eq:scac_1_7} and the dynamic constraints \eqref{eq:scac_13}, where the stability constraints are also written in an implicit form.
By expressing the generator dynamics and the stability constraints in the same implicit form as the static equality and inequality constraints \eqref{eq:scac_1_7}, the system dynamics can be incorporated into the static constraints of model \ref{model:stability_constrained_ac_opf}, and integrated seamlessly 
into the optimization process. This unified framework simplifies handling both the static and dynamic requirements of the problem.
Importantly, the stability constraints \eqref{eq: stab-limit} are estimated as
\begin{equation}
    \hat{\delta}^g(t) - \delta^{\max} \leq 0 \; \; \forall g \in \pazocal{G},
    \label{eq: est_stab_limit}
\end{equation}
which enables, together via \eqref{eq:LD_loss} and \eqref{eq:LD_loss_constr}, end-to-end training of the DynOPF-Net model because of the \emph{differentiable} nature of the generator dynamic predictor $\mathcal{N}^g_\phi$. 
At iteration $k+1$, finding the optimal DynOPF-Net parameters requires solving 
\[
(\psi,\phi)^{k+1} = \arg\min_{\psi, \phi} \mathbb{E}_{\left(\bm{S}^d, \bm{y}^\star\right) \sim \pazocal{D}}
\left[ \pazocal{L}^{\text{DynOPF}} \left(\mathcal{D}_{\psi^k}^{\bm{\lambda}^k}(\bm{S}^d), \bm{y}^\star, \right. \right.
\left. \left. \mathcal{N}_{\phi^k}^{\bm{\lambda}^k}(\hat{\bm{x}}(0), \hat{V}_g)) \right) \right],
\]
where $\mathcal{D}_{\psi}^{\lambda^k}$, $\mathcal{N}_{\phi}^{\lambda^k}$ denotes the DynOPF-Net proxy optimization model $\mathcal{D}_{\psi^k}$ and the array of NODE models $\mathcal{N}^g_{\phi}, \forall g \in \pazocal{G}$ at iteration $k$, with $\bm \lambda^k = [ {\lambda}^k_{h_1}, \ldots, {\lambda}^k_{h_{n_{eq}}}, {\lambda}^k_{u_1}, \ldots, {\lambda}^k_{u_{n_{ineq}}}]^T$. This step is approximated using a stochastic gradient descent method
\begin{equation}
\psi^{k+1} = \psi^k - \eta \nabla_{\psi} \pazocal{L}^{\text{DynOPF}}\left(\mathcal{D}_{\psi^k}^{\bm{\lambda}^k}(\bm{S}^d), \bm{y}^\star, \hat{\bm x}(t) \right) \notag
\end{equation}
\begin{equation}
\phi^{k+1} = \phi^k - \eta \nabla_{\phi} \pazocal{L}^{\text{DynOPF}}\left(\hat{\bm y},
\bm{y}^\star, \mathcal{N}_{\phi^k}^{\bm \lambda^k}(\hat{\bm{x}}(0), \hat{V}_g)\right), \notag
\end{equation}
where $\eta$ denotes the learning rate and $\nabla_{\psi (\phi)} \pazocal{L}^{\text{DynOPF}}$ represents the gradient of the loss function $\pazocal{L}^{\text{DynOPF}}$ with respect to the parameters $\psi (\phi)$ at the current iteration. 
Note this step does not retrain $\mathcal{D}_{\psi}$, $\mathcal{N}_{\phi}$ from scratch, but uses a hot start for the weights $\psi$, $\phi$.
Finally, the Lagrange multipliers are updated as 

\begin{equation}
 {\lambda}_{h_j}^{k+1} = {\lambda}_{h_j}^{k} + \rho \; \nu \left( h_j \left( \hat{\bm y}, \hat{\bm{x}}(t) \right) \right) \; \;  j=1,\ldots,n_{eq} \notag
\end{equation}
\begin{equation}
 {\lambda}_{u_l}^{k+1} = {\lambda}_{u_l}^{k}+ \rho \; \nu \left( u_l \left( \hat{\bm y}, \hat{\bm{x}}(t)) \right) \right) \; \; l=1,\ldots,n_{ineq} \; , \notag
\end{equation}


where $\rho$ denotes the Lagrangian step size. The
DynOPF-Net training scheme is presented in Algorithm \ref{alg:LD}. It takes as input the training dataset $\pazocal{D}$, learning rate $\eta>0$, and step size $\rho>0$. The Lagrange multipliers $\bm{\lambda}$ are initialized in line $2$.
As shown in Figure~\ref{fig: DynOPF-Net}, for epoch $k$ and each sample $i$, given 
$\bm{S}^{d,i}$ 
(line $4$), the DynOPF-Net's proxy optimization model $\mathcal{D}_{\psi^k}$ computes an estimate of the OPF variables $\hat{\bm y}^i$. Given these estimates, 
for each generator, $g \in \pazocal{G}$, the initial values of the state variables and EMF are computed according to Eq. \eqref{eq: init_delta},  \eqref{eq: init_EMF}, \eqref{eq: init_omega}
(line $7$). For each generator $g \in \pazocal{G}$, the estimated initial values $\hat{\bm x}^{g,i}(0)$ are input to NODE $\mathcal{N}^g_{\phi^k}$, which computes an estimate of the generator state variables $\hat{\bm x}^{g,i}(t)$ (line $8$), based on which the violation of each stability constraint (line $9$) is computed. The OPF variables' prediction error 
and the total constraint violations 
are then used to compute the loss function $\pazocal{L}^\text{DynOPF}$ (line $10$) \eqref{eq:LD_loss} as specified by \eqref{eq:LD_loss_pred}-\eqref{eq: est_stab_limit}, using predicted values $\hat{\bm y}$, $\hat{\bm x}(t)$ and multipliers $\bm \lambda^k$ at the current epoch $k$. The weights $\psi, \phi$ of the DynOPF-Net model are then updated using stochastic gradient descent (lines $11$).
Finally, at the end of the epoch, the multipliers are updated based on the respective constraint violations (line $12$). 


\begin{algorithm}
\caption{DynOPF-Net for stability-constrained OPF}
\label{alg:LD}
\begin{algorithmic}[1]
\small
\State \textbf{Input:} Dataset ${\pazocal D} = \{\left( \bm{S}^{d,i}, \bm{y}^{\star,i}\right)\}_{i=1}^{n_\text{obs}}$; optimizer method, learning rate $\eta$ and Lagrangian step size $\rho$.
\State Initialize Lagrange multipliers $\bm{\lambda}_{h}^{0} = 0$, $\bm{\lambda}_{u}^{0} = 0$.
\State \textbf{For} {each epoch $k = 0, 1, 2, \dots$}
\State \ \ \ \textbf{For} {each $\left( \bm{S}^{d,i}, \bm{y}^{\star,i}\right) \in \pazocal{D}$}
\State \ \ \ \ \ \ $\hat{\bm y}^i \gets \mathcal{D}_{\psi^k}(\bm{S}^{d,i})$.
\State \ \ \ \ \ \ \textbf{For} {each generator $g \in \pazocal{G}$}
\State \ \ \ \ \ \ \ \ \ Compute $\hat{\bm x}^{g,i}(0)$ using \eqref{eq: init_delta}-\eqref{eq: init_omega}.
\State \ \ \ \ \ \ \ \ \  $\hat{\bm x}^{g,i}(t) \gets \mathcal{N}^g_{\phi^k}(\hat{\bm x}^{g,i}(0), \hat{V}_g^i)$.
\State \ \ \ \ \ \ \ \ \ Compute $\max(0, \hat{\delta}^{g,i}(t)-\delta^{\max})$.
\State \ \ \ \ \ \  {$\pazocal{L}^{\text{DynOPF}}(\hat{\bm y}^{i}, \bm y^{\star,i}, \hat{\bm{x}}^{i}(t)) \gets \pazocal{L}_p(\hat{\bm y}^{i}, \bm{y}^{\star,i}) + \pazocal{L}_c(\hat{\bm y}^{i}, \hat{\bm{x}}^{i}(t)).$} 
\State \ \ \ \ \ \ Update DynOPF-Net model $\mathcal{D}_{\psi^k}$, $\mathcal{N}_{\phi^k}$ parameters
        \[
        \psi^{k+1} \gets \psi^k - \eta \nabla_{\psi} \pazocal{L^{\text{DynOPF}}}(\mathcal{D}^{\bm \lambda^k}_{\psi^k}(\bm 
        {S}^{d,i}), \bm y^{\star,i}, \hat{\bm x}^{i}(t))
        \]
        \[
        \phi^{k+1} \gets \phi^k - \eta \nabla_{\phi} \pazocal{L^{\text{DynOPF}}}
        (\hat{\bm y}^i, \bm y^{\star,i},
        \mathcal{N}_{\phi^k}^{\bm \lambda^k}(\hat{\bm{x}}^{g,i}(0), \hat{V}_g^i)).
        \]    
\State \ \ \ Update Lagrange multipliers
        \[
        {\lambda}_{h_j}^{k+1} \gets {\lambda}_{h_j}^{k} + \rho \; \nu \left( h_j \left(  \hat{\bm y}, \hat{\bm{x}}(t) \right) \right) \; \;  j=1,\ldots,n_{eq},
        \]
        \[
        {\lambda}_{u_l}^{k+1} \gets {\lambda}_{u_l}^{k}+ \rho \; \nu\left( u_l \left( \hat{\bm y}, \hat{\bm{x}}(t) \right) \right) \; \; l=1,\ldots,n_{ineq}.
        \]
    \EndFor
\EndFor
\end{algorithmic}
\end{algorithm}


\section{Experimental Setting} \label{sec: experimental setting}
The effectiveness of DynOPF is tested on two power networks of different sizes and complexity: the WSCC 9-bus system
, and the IEEE 57-bus system \cite{babaeinejadsarookolaee2021power}. Our approach is benchmarked against $3$ widely adopted LtO methods approximating the AC-OPF problem \ref{model:ac_opf} solutions $\bm y = (\bm{S}^r, \bm V)$ given $\bm{S^d}$.
 With reference to \eqref{eq:LD_loss}, \eqref{eq:LD_loss_pred}, \eqref{eq:LD_loss_constr}, they use the following loss functions for model training.
\begin{itemize}
    \item Zamzam et.~al \cite{Zamzam_learn_19} uses a loss function:
    \begin{equation*}
    \begin{cases}
    \pazocal{L}_p(\bm{y}, \hat{\bm{y}}) = \|\hat{\bm y}-\bm y ^\star\|^2 \\
    \pazocal{L}_{c}(\hat{\bm y}, \hat{\bm{x}}(t)) = 0,
    \end{cases}
    \end{equation*}
    which minimizes the mean squared error (\textbf{MSE}) between the predicted solution and its corresponding label.
    \item \textbf{Lagrangian-Dual} \cite{fioretto2020predicting} uses a Lagrangian loss function:
    \begin{equation*}
    \begin{cases}
    \pazocal{L}_p(\bm{y}, \hat{\bm{y}}) = \|\hat{\bm y}-\bm y ^\star\|^2 \\
    \begin{aligned}
    \pazocal{L}_{c}(\hat{\bm y}, \hat{\bm{x}}(t)) = \sum_{j=1}^{n^\prime_{eq}} {\lambda}_{h_j} \nu \left( h_j(\hat{\bm y}, \hat{\bm{x}}(t)) \right) + \sum_{l=1}^{n^\prime_{ineq}}{\lambda}_{u_l} \nu \left( u_l(\hat{\bm y}, \hat{\bm{x}}(t) ) \right),
    \end{aligned}
    \end{cases}
    \end{equation*}
    where  $n^\prime_{eq}$, $n^\prime_{ineq}$ denotes the number of static equality and inequality constraints of model \ref{model:stability_constrained_ac_opf}, specified by functions $h_j, \; j=1,\ldots,n^\prime_{eq}$ and $u_l, \; l=1,\ldots,n^\prime_{ineq}$ \eqref{eq:scac_1_7}. 
    \item \textbf{DC3} \cite{donti2020dc3} uses the loss function: 
    \begin{equation*}
    \begin{cases}
    \pazocal{L}_p(\bm{y}, \hat{\bm{y}}) = f(\hat{\bm{y}}) \\
    \begin{aligned}
    \pazocal{L}_{c}(\hat{\bm y}, \hat{\bm{x}}(t)) = {\lambda}_{h} \sum_{j=1}^{n^\prime_{eq}}  \| h_j(\hat{\bm y}, \hat{\bm{x}}(t)) \|^2 + {\lambda}_{u}\sum_{l=1}^{n^\prime_{ineq}} \max(0, u_l(\hat{\bm y}, \hat{\bm{x}}(t) )^2),
    \end{aligned}
    \end{cases}
    \end{equation*}
    where $f$ is objective function \eqref{scac_obj}.
    This method relies on a completion-correction technique to enforce constraints \eqref{eq:scac_1_7} satisfaction of Model \ref{model:stability_constrained_ac_opf}, 
    in self-supervised training. 
\end{itemize}
The rest of this section describes the training setting. 


\subsubsection*{Practical considerations}
since the 
the generator model \eqref{eq: generator dynamics} is known,
to obtain accurate estimates of the state variables 
each neural ODE model can be hot-started 
and then integrated within the DynOPF-Net model as detailed in Section \ref{sec:DynOPF-Net}.
\begin{figure*}[t!]
    \centering
    \begin{subfigure}[b]{0.48\textwidth}
        \centering
        \includegraphics[height=5cm]{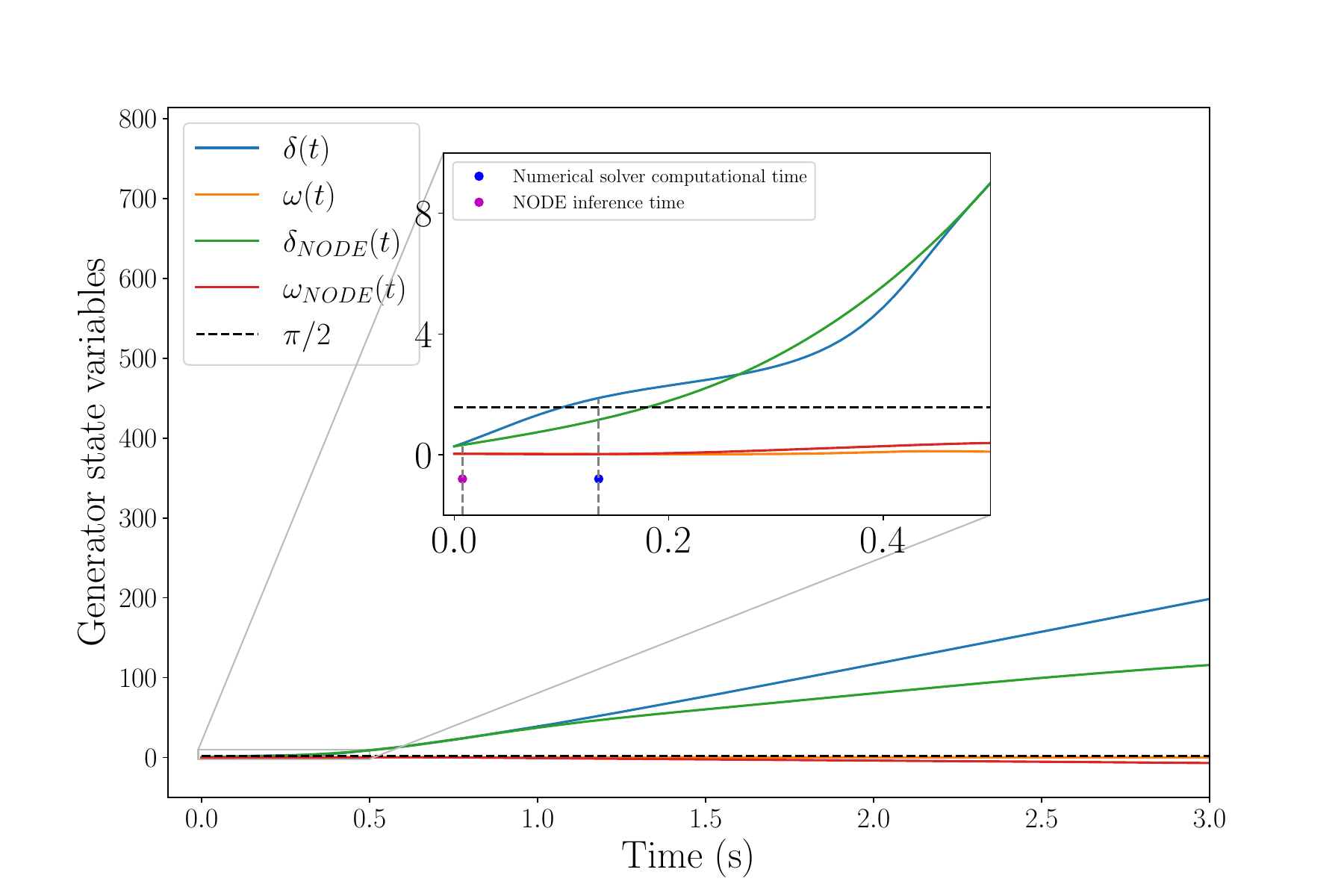}  
        \caption{Numerical and NODE estimated solutions of system \eqref{eq: generator dynamics} in unstable conditions. The NODE model detects instability conditions much faster than a numerical solver and before the physical system transitions into an unstable state.}
        \label{fig:fast_unstable_dynamics}
    \end{subfigure}
    \hfill
    \begin{subfigure}[b]{0.48\textwidth}
        \centering
        \includegraphics[height=5cm]{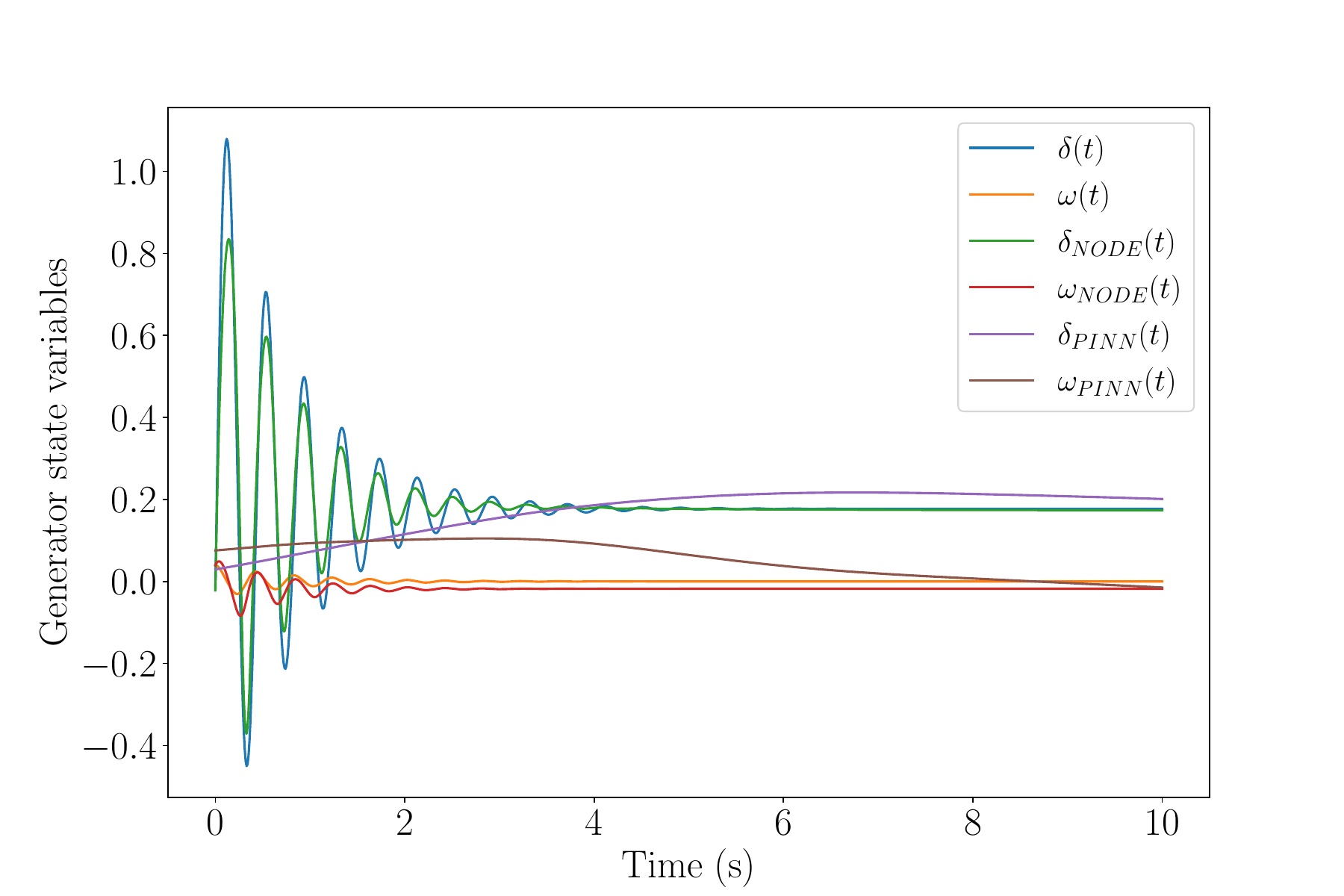}  
        \caption{Numerical, PINN, and NODE estimated solutions of system (\ref{eq: generator dynamics}) in stable conditions.}
        \vspace{20pt}
        \label{fig:stable_traj}
    \end{subfigure}
    \caption{Comparison of solutions in unstable and stable conditions using a numerical ODE solver, NODE, and PINN model.}
\end{figure*}
\paragraph{Dataset creation and training setting}
Based on the motivations above, each NODE model $\mathcal{N}_{\phi}^g$ is trained in a supervised fashion as described in Section \ref{sec: NODE}. For each generator $g \in \pazocal{G}$, the dataset $\pazocal{D}^g$ used for training $\mathcal{N}_{\phi}^g$ consists of 
distinct time series, each representing the solution of \eqref{eq: generator dynamics} given a different instance of the generator model. Some operational set points yield stable conditions, while others are unstable. 
The NODE models are trained on a dataset $\pazocal{D}^g=\{ \left((\bm{x}^{g,i}(0), V^i_g), (\bm{x}^{g,i}(t),V^i_g)\right) \}_{i=1}^{10,000}$, where
$(\bm{x}^{g}(0), V_g)$ represents the input to the dynamic predictor $\mathcal{N}_{\phi}^g$ 
as discussed in \ref{sec: NODE}, while $\bm{x}^g(t) = (\delta^g(t), \omega^g(t), V_g)$ is the corresponding target and solution of the augmented generator model (\ref{eq: NODE dynamics predictors}).
For each $\bm{x}^g(0), V_g$, each OPF variable 
$S^r_g, V_g$ 
is sampled from a uniform distribution in which lower and upper bounds are given by the corresponding operational limits \eqref{eq:ac_1}, \eqref{eq:ac_6} and \eqref{eq:ac_2}, \eqref{eq:ac_5}. Note that this sampling scheme spans the full range of \emph{feasible} OPF variables; since the NODEs' inputs are provided by $\mathcal{D}_{\psi^k}$; this approach is robust with respect to its small prediction error, as verified in practice. 
Given the values of $|V_g|, \theta_g$, the initial state variables values $\delta^g(0)$ and $\omega^g(0)$ are obtained from \eqref{eq: init_delta}-\eqref{eq: init_omega}. The parameters of generator model \eqref{eq: generator dynamics}, such as the damping coefficient $d^g$ and inertia constant $m^g$, are adopted from \cite{article_parameter}.
Given initial condition $\bm{x}^g(0)$, the corresponding solution of \eqref{eq: generator dynamics} $\bm{x}^g(t)$ is generated by adopting a numerical ODE algorithm named
$Dopri5$ \cite{liu2010comparison}, an adaptive Runge-Kutta method of order
$5$. The simulation time is set to $3$ seconds (s) and the initial $\Delta_t$ is set at $0.001$ (s). 
Each dataset $\pazocal{D}^g$ is divided into training, validation, and test sets with a $80/10/10$ split.

On both the IEEE 57 and WSCC 9-bus systems, DynOPF-Net and each baseline LtO model is trained on a dataset ${\pazocal D} = \{\left( \bm{S}^{d,i},
\bm{y}^{\star,i}\right)\}_{i=1}^{10,000}$.
Each input 
$\bm{S}^d$ represents a load vector, while the corresponding target is the associated optimal generator set-points $\bm{y}^\star=\bm{(S^r, V)}$
of Model \ref{model:stability_constrained_ac_opf}, with the assumption that the generators are in steady-state.
Similarly to \cite{fioretto2020predicting}, the load demands vector $S_i^d$ 
is generated by applying to each nominal load $S_i^d, \; i \in \pazocal{N}$, a uniform random perturbation which results in load demands set to $\pm20\%$ $S_i^d$. The dataset only reports feasible AC-OPF instances. Thus, each training data point represents a valid snapshot of the AC-OPF problem, consisting of a load profile along with the corresponding optimal generator set points.
The AC-OPF models are implemented with MATPOWER and solved with IPOPT, a numerical implementation of the interior point method \cite{Wchter2006OnTI}. 
For each test case, the dataset $\pazocal{D}$ is divided into training, validation, and test sets on a $80/10/10$ split.
\begin{table}[h]
\centering
\caption{Average and standard deviation of computational times for numerical solvers vs.~neural ODE inference times}
\label{tab: comptime_methods}
\begin{tabular}{l c c}
\toprule
Method & Numerical Solver & NODE \\ \midrule
Dopri5 (default) & $0.135\pm0.015$ (sec) &  \bm{$0.008\pm0$} (sec)\\
Bosh3 &  $0.446\pm0.039$ (sec) & \bm{$0.017\pm0$} (sec)\\
\bottomrule
\end{tabular}
\end{table}

\begin{table*}[ht]
\centering
\caption{Average and standard deviation of MSE, constraint violations, and steady-state gap for the WSCC $9$-bus system across different approaches based on $40$ trials.}
\label{tab:9-bus comp}
\begin{adjustbox}{width=\textwidth,center}
\begin{tabular}{l c c c c c c c c}
\toprule
\multirow{2}{*}{Method} & \multicolumn{5}{c|}{Steady-State Metrics} & \multicolumn{3}{c}{Dynamic Metrics} \\
\cmidrule(lr{0.5em}){2-9}
& \makecell{MSE ($\bm{p^r}$) \\$\times 10^{-3}$} & \makecell{MSE ($\bm{q^r}$) \\$\times 10^{-3}$} & \makecell{MSE ($|\bm{V}|$) \\$\times 10^{-4}$}& \makecell{MSE ($\bm{\theta}$) \\$\times 10^{-4}$} & \makecell{Optimality Gap \% \\ $\times 10^{-1}$} & \makecell{Flow Vio. \eqref{eq:ac_3}\\ $\times 10^{-3}$} & \makecell{Boundary Vio.  \eqref{eq:ac_1}-\eqref{eq:ac_5} \\$\times 10^{-4}$}& \makecell{\textbf{Stability Vio.} \eqref{eq:scac_13} \\  \bm{$\times 10^{1}$}} \\ \midrule

MSE & $1.90 \pm 0.272 $ & $1.65 \pm 1.018$ & $0.32 \pm 0.153$ & $0.43 \pm 0.149$ & $1.29 \pm 0.008$ & $10.45 \pm 2.183$ & $9.72 \pm 4.930$  & \bm{$2.26 \pm 0.189$} \\

DC3 & $1.86 \pm 0.217$ & $1.56 \pm 0.326$ & $0.26 \pm 0.195$ & $0.48 \pm 0.343$ & $1.17 \pm 0.006$ & $0.00$ & $0.00$ & \bm{$2.45 \pm0.205$}  \\

LD  & $1.77 \pm 0.163 $ & $1.72 \pm 0.248$ & $0.16 \pm 0.099$ & $0.55 \pm 0.051$ & $1.12 \pm 0.008$ & $7.19 \pm 0.425$ & $0.00$ & \bm{$2.13 \pm 0.175$}  \\

DynOPF-Net & $2.41 \pm 0.253$ & $3.18 \pm 1.273$ & $2.55 \pm 0.354$ & $3.82 \pm 0.924$ & $1.32 \pm 0.032$ &  $8.32 \pm 0.596$ & $0.41 \pm 0.243$ & \bm{$0.00$} \\
\bottomrule
\end{tabular}
\end{adjustbox}
\end{table*}

\begin{table*}[ht]
\centering
\caption{Average and standard deviation of MSE, constraint violations, and steady-state gap for the IEEE $57$-bus system across different approaches based on $40$ trials.}
\label{tab:57-bus comp}
\begin{adjustbox}{width=\textwidth,center}
\begin{tabular}{l c c c c c c c c}
\toprule
\multirow{2}{*}{Method} & \multicolumn{5}{c|}{Steady-State Metrics} & \multicolumn{3}{c}{Dynamic Metrics} \\
\cmidrule(lr{0.5em}){2-9}
&  \makecell{MSE ($\bm{p^r}$) \\$\times 10^{-3}$} & \makecell{MSE ($\bm{q^r}$) \\$\times 10^{-3}$} & \makecell{MSE ($|\bm{V}|$) \\$\times 10^{-3}$}& \makecell{MSE ($\bm{\theta}$) \\$\times 10^{-3}$} & 
\makecell{Optimality Gap \%\\ $\times 10^{-1}$} &  \makecell{Flow Vio. \eqref{eq:ac_3}\\ $\times 10^{-3}$} 
& \makecell{Boundary Vio. \eqref{eq:ac_1}-\eqref{eq:ac_5} \\$\times 10^{-4}$}& 
\makecell{\textbf{Stability Vio.} \eqref{eq:scac_13} \\ \bm{$\times 10^{1}$}}\\ \midrule
MSE & $3.48 \pm 0.321$ & $3.86 \pm 1.512$  & $0.77 \pm 0.148$ & $1.42 \pm 0.237$ & $1.66 \pm 0.243$ & $12.65 \pm 2.281$ & $6.44 \pm 1.434$ & \bm{$2.33 \pm 0.206$}  \\

DC3 & $3.31 \pm 0.579$ & $6.74 \pm 0.580$ & $0.51 \pm 0.078$ & $0.64 \pm 0.081$ & $1.64 \pm 0.049$ & $0.00$ & $0.00$ & \bm{$2.86 \pm 0.232$}  \\

LD & $3.97 \pm 0.279$ & $3.52 \pm 2.427$ & $0.34 \pm 0.012$ & $0.95 \pm 0.054$ & $1.68 \pm 0.125$ & $6.23 \pm 0.125$ & $0.00$ & \bm{$2.31 \pm 0.219$}  \\

DynOPF-Net & $5.05 \pm 0.175$ & $7.42 \pm 1.482$ & $2.99 \pm 0.214$ & $4.43 \pm 0.673$ & $2.26 \pm 0.180$ & $9.15 \pm 0.442$ & $0.25 \pm 0.172$ & \bm{$0.00$} \\
\bottomrule
\end{tabular}
\end{adjustbox}
\end{table*} 

\section{Experimental Results} \label{sec: experimental result}
This section presents the results of each benchmark system of DynOPF-Net and each baseline LtO method. 
The experiments focus on two main aspects: {\bf (1)} assessing the effectiveness of NODEs and PINNs in capturing generator dynamics and comparing their precision and computational efficiency to a numerical ODE solver; {\bf (2)} comparing DynOPF-Net with LtO methods that capture only the steady-state AC-OPF problem, focusing on the stability constraint violations.
Unless otherwise stated, results are reported as an average of 40 independent runs on a subset of dataset $\pazocal{D}$ where DynOPF-Net and each LtO model is \emph{not} trained on, which we refer to as the test set. Specifically, we report:
\begin{itemize}[leftmargin=*,topsep=0pt]
    \item The inference time (measured in seconds) of NODEs which we compare to the computational time of a traditional numerical ODE solver and the precision of state variables' estimate (measured as the percentage relative $\ell^2$ error) of NODEs and PINNs, a different learning based approach discussed in Section \ref{subsec: dyn_forcasting}.
    \item The average constraint violations (measured as 
    $\frac{1}{n_{test}} \sum_{i=1}^{n_{test}}| h_j(\hat{\bm y}^i, \hat{\bm{x}}^i(t)) | $ for the $j$-th equality and $\frac{1}{n_{test}}\sum_{i=1}^{n_{test}}\max(0, u_l(\hat{\bm y}^i, \hat{\bm{x}}^i(t) ))$ for the $l$-th inequality, where $n_{test}$ is the test set size),  incurred by DynOPF-Net and each baseline method approximation of ${\bm y}$.
    \item The average percentage \emph{steady-state} gap incurred by DynOPF-Net and the baselines LtO predictions $\hat{\bm y}$, and measured as $\frac{|f(\bm{y}^\star(\bm{S}^d)-f(\hat{\bm{y}}(\bm{S}^d))|}{|f(\bm{y}^\star(\bm{S}^d))|} *100$, where $f$ is objective \eqref{scac_obj} and prediction error (MSE) $\| \bm y^\star - \hat{\bm y}\|^2$ of the OPF variables with respect to optimal (steady-state) generators set-points values.
    {The steady-state assumption is crucial for evaluating how closely each solution approximates the AC-OPF optimal results, though it \emph{does not reflect the stability-constrained AC-OPF problem results} which is the main focus of this paper. Given the non-linearity of both the dynamics and optimization in the stability-constrained AC-OPF, computing exact optimal decisions $\bm{y}^\star$ with traditional methods is unfeasible. Consequently, while our method may achieve slightly higher steady-state gaps or prediction errors, \emph{these should not be interpreted in the context of the stability-constrained AC-OPF problem} and the key desiderata and goal is that of producing solutions with low stability violations.}
    \item The inference time (measured in seconds) of DynOPF-Net and each LtO model to generate $\hat{\bm y}$.
\end{itemize}

\subsection{Dynamic Forecasting}\label{subsec: dyn_forcasting}
\paragraph{Runtime comparison between NODEs and a traditional ODE solver}
Here the goal is to evaluate the NODEs' inference time to produce the generator state variables' estimates and to compare it with the computational time of a traditional ODE solver.
Given the synchronous generator model described by 
(\ref{eq: generator dynamics}), a numerical ODE solver could be adopted to
determine the evolution in time of the state variables $\delta^g(t)$ and $\omega^g(t)$. However, in case of unstable conditions, the system dynamics can be as rapid as, or even surpass, the time required for computing the ODE solution with a numerical solver.
This situation is depicted in Figure~\ref{fig:fast_unstable_dynamics} where unstable conditions arise before a numerical solution to the system of differential equations~\eqref{eq: generator dynamics} is computed. Conversely, the neural ODE model $\mathcal{N}^g_{\phi}$ is capable of detecting unstable conditions before the system transitions into an unstable state, while also providing a good approximation of the solution. As anticipated in Section \ref{sec: NODE}, this speed advantage arises from the vector field approximation of (\ref{eq: generator dynamics}), enabling quicker computation of update steps in a numerical ODE solver. 
Table~\ref{tab: comptime_methods}
reports the average and standard deviation of computational time, for numerical solvers, and inference time, for neural ODEs, given $2$ different numerical algorithms. \emph{Neural ODE models are, on average, about $20$ times faster than a numerical solver which uses the governing equations of~\eqref{eq: generator dynamics}. 
This aspect makes neural ODE models natural candidates as dynamic predictors for the generator model in real-time applications.} 

\paragraph{Comparison between NODEs and PINNs.}
Here the goal is to assess the precision of the NODEs' estimate of the generator state variables and to compare them with PINNs \cite{Misyris2019PhysicsInformedNN}. PINNs are ML-based models that incorporates known physical laws 
into the learning process. Instead of relying solely on data, PINNs use physics-based constraints to guide the training, ensuring that the model’s predictions are consistent with the underlying scientific principles.
Figure~\ref{fig:stable_traj} shows the NODE and PINNs' state variables estimates in case of stable conditions.
While a NODE model produces highly accurate state variables' predictions, 
a PINN model trained on the same dataset $\pazocal{D}^g$ but affected by a generalization bias, 
is incapable of capturing the generator dynamics across different instances of~\eqref{eq: generator dynamics} and produces a poor state variables' estimate. 
Specifically, the percentage $\ell^2$ error between the numerical ODE solver solutions $\delta(t), \omega(t)$ and the NODE predictions $\delta_{\text{NODE}}(t), \omega_{\text{NODE}}(t)$ is $5.17\%$, while 
for the PINN predictions $\delta_{\text{PINN}}(t), \omega_{\text{PINN}}(t)$ is significantly higher at $69.45 \%$.

\begin{figure*}[t!]
    \centering
    \begin{subfigure}[b]{0.48\textwidth}
        \centering
        \includegraphics[height=4.2cm]{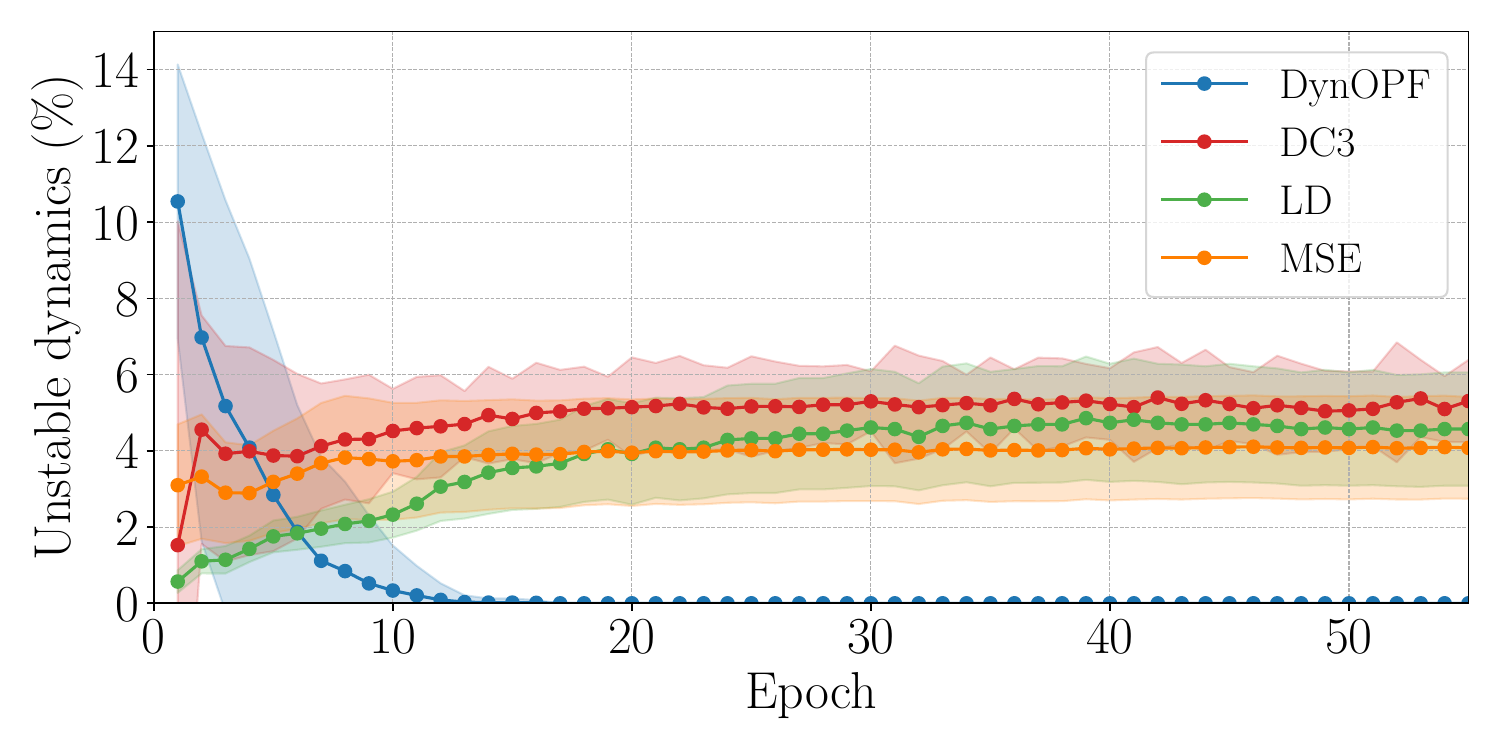}  
        \caption{WSCC $9$-bus system - Percentage of unstable solutions at training time for different methods.}
    \label{fig:9Bus_unstab_traj_LD}
    \end{subfigure}
    \hfill
    \begin{subfigure}[b]{0.48\textwidth}
        \centering
        \includegraphics[height=4.2cm]{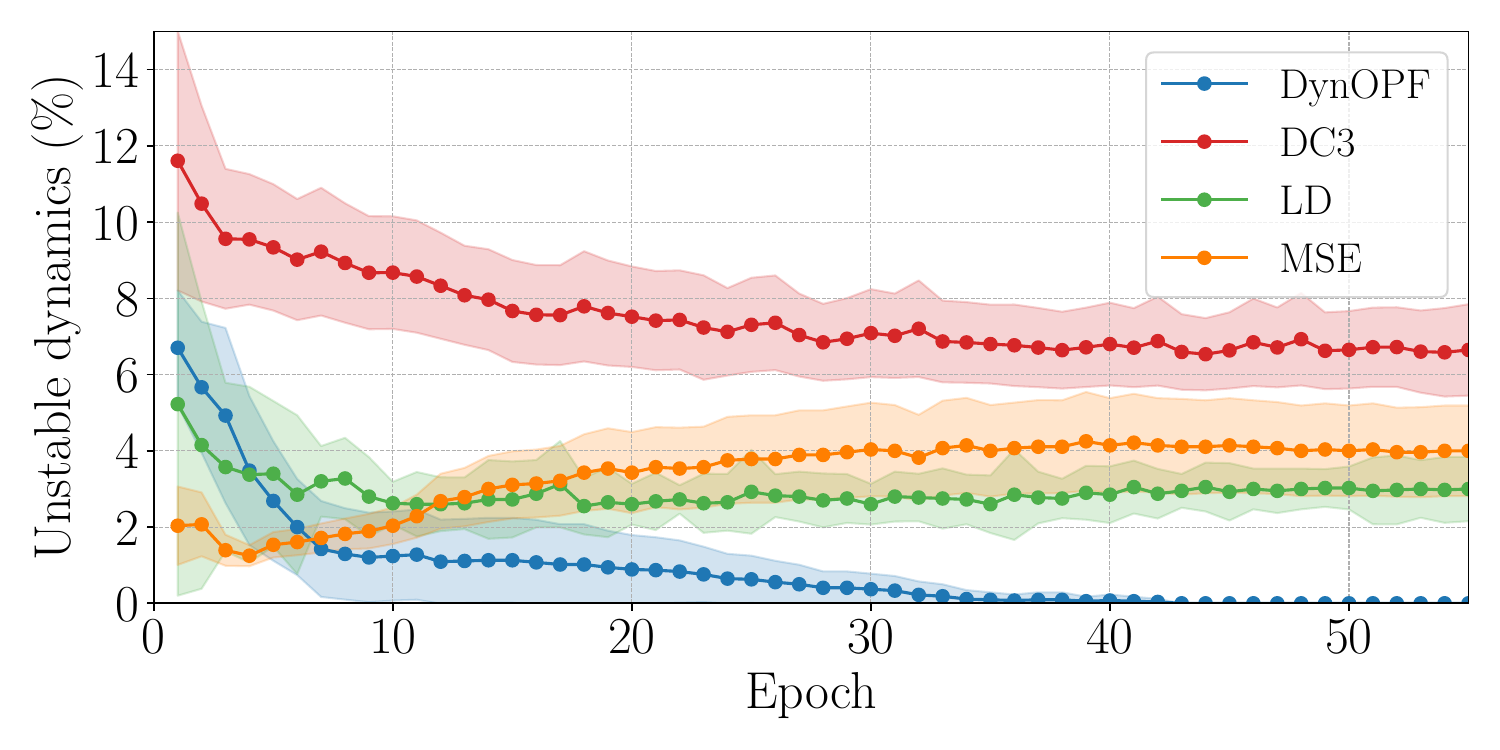}  
        \caption{IEEE $57$-bus system - Percentage of unstable solutions at training time for different methods.}
        \label{fig:57Bus_unstab_traj_LD}
    \end{subfigure}
\end{figure*}






\subsection{Steady-state Prediction Errors and Constraint Violations}
Next, we investigate constraint violations, with a focus on the stability constraint violations~\eqref{eq:scac_13} and how they relate with \emph{steady-state} prediction errors. 
Tables~\ref{tab:9-bus comp} and \ref{tab:57-bus comp}
report the average and standard deviation of predicted set-point errors (MSE), constraint violations and steady-state gap (which will be discussed in the next subsection) on the test set, for each method and benchmark system.
First note that, for each test case, the tables report comparable prediction errors and static constraint violations~\eqref{eq:scac_1_7}, across different methods. DC3 achieves no static (flow and boundary) constraint violations, being designed to ensure feasibility during training.
However, all baseline methods systematically fail to satisfy stability requirements~\eqref{eq:scac_13}. By integrating the generator dynamics within training, DynOPF-Net learns to meet the stability requirements in the first stage of training, as seen in Figures~\ref{fig:9Bus_unstab_traj_LD} and \ref{fig:57Bus_unstab_traj_LD}, achieving compliance with the dynamic requirements around epochs $20$ and $50$, respectively. Test results in Tables~\ref{tab:9-bus comp} and \ref{tab:57-bus comp} show the stability constraint violations align with training.

Figures~\ref{fig:9Bus_unstab_traj_LD} and \ref{fig:57Bus_unstab_traj_LD} show the percentage of solutions violating the stability constraints in the first $50$ epochs of training.
Figure~\ref{fig:9Bus_unstab_traj_LD}, pertaining to the WSCC 9-bus system, shows that DynOPF-Net learns rapidly to meet the dynamic constraints, which violations approach zero level after epoch $10$ of training. In contrast, all the baseline methods produce between 
$4\%$ and $6\%$ of unstable solutions throughout training. Figure~\ref{fig:57Bus_unstab_traj_LD} shows a similar scenario for the IEEE 57-bus system. DynOPF-Net learns to address the dynamic requirements and achieves no stability constraint violations at training time after epoch $40$. Conversely, the baseline methods 
produce between
$4\%$ and $7\%$ of unstable solutions during training. The convergence rate of DynOPF-Net is slightly slower than on the WSCC-9 bus system, likely due to the complexity of the test case. \emph{These results provide strong empirical evidence of the importance to integrate dynamic requirements into the AC-OPF problem.}
Our approach learns to modify potentially unstable set points, at a cost of only slightly higher steady-state MSE than the baseline approaches. Note in particular the MSE of $|\bm{V}|$ and $\bm{\theta}$; these variables directly affect the generator dynamics in \eqref{eq: generator dynamics}, and thus their modification is necessary to ensure stability constraints satisfaction. 
In contrast, all the baseline methods achieve slightly smaller steady-state prediction error 
than DynOPF-Net, as they solve the steady-state AC-OPF problem \ref{model:ac_opf} ignoring the generator dynamics, and produce significant violations of the stability constraints. \emph{These results indicate that the OPF voltage setpoints have high impact on the generator dynamics.} 

\subsection{Steady-state Gaps} 
This section discusses the suboptimality of the estimated solution $\bm{\hat{y}}$ with respect to the ground truth $\bm{y}^\star$, given loads 
$\bm{S}^d$ and measured in terms of objective value~\eqref{scac_obj}, of DynOPN-Net and each baseline method. 
{Note that the steady-state gap is measured with respect to the optimal solution $\bm{y}^\star$ of the stability-constrained AC-OPF Model \ref{model:stability_constrained_ac_opf}, with the assumptions that the generators are in steady-state conditions, and thus does not measure the stability-constrained AC-OPF optimality gap, which is unattainable in the setting considered.}
Nonetheless, it provides valuable insights on the decision quality of each LtO method and DynOPF-Net for solving Problem \ref{model:stability_constrained_ac_opf}.
Tables~\ref{tab:9-bus comp} and \ref{tab:57-bus comp} report the steady-state gaps on the WSCC-9 and IEEE-57 bus systems, respectively.
The tables report that 
all methods achieve comparable gaps in each test case. This is intuitive, since  objective \eqref{scac_obj} depends solely on the power generated $\bm{p}^r$, and all methods produce similar $\bm{p}^r$'s prediction error, as displayed in Tables~\ref{tab:9-bus comp} and \ref{tab:57-bus comp}.
DynOPF-Net produces average percentage steady-state gaps of $1.32 \times 10^{-1} \%$ and $2.26 \times 10^{-1} \%$ for the WSCC 9 and the IEEE 57-bus system while preserving system stability, that are comparable with the best gaps - LD with $1.12 \times 10^{-1} \%$ and DC3 with $1.64 \times 10^{-1} \%$ - which often violates stability constraints.
{The higher objective costs observed with DynOPF-Net is intuitively attributed to a restricted feasible space due to the integration of generator stability constraints \eqref{eq:scac_13} within the AC OPF model \ref{model:ac_opf}.}

\begin{table}[h!]
\centering
\caption{Average and standard deviation of inference times for different OPF learning approaches} 
\label{tab: inference_time}
\begin{tabular}{l c c}
\toprule
Method & WSCC 9-bus & IEEE 57-bus \\ \midrule
DynOPF-Net &  $0.001\pm 0.00$ (sec) & $0.009\pm 0.00$ (sec)\\
DC3  & $0.025\pm 0.00$ (sec) &  $0.089\pm 0.00$ (sec)\\
LD  & $0.000\pm 0.00$ (sec) &  $0.001\pm 0.00$ (sec)\\
MSE  & $0.000\pm 0.00$ (sec) &  $0.001\pm 0.00$ (sec)\\
\bottomrule
\end{tabular}\vspace{-3mm}
\end{table}

\subsection{Inference Time}
Finally, we evaluate the average inference time of DynOPF-Net and each baseline LtO model. 
Table~\ref{tab: inference_time} shows the inference time of each LtO method on each test case.
On average, DynOPF produces near-optimal and stable solutions in $1$ (ms) and $9$ (ms) for the WSCC-9 bus and IEEE 57-bus, respectively, which is slightly slower than the MSE and LD approaches, and about $15 \times$ faster than the DC3 method. 
These results suggest that there is also a tradeoff between compliance with dynamic requirements and inference time.
DC3 achieves the highest inference time, due to its correction and completion procedure, which requires solving a nonlinear system of equations and the Jacobian matrix computation. 
While DynOPF-Net is already applicable for real-time applications, its efficiency could be improved by computing the state variables in parallel, since each dynamic predictor is independent. \emph{This aspect makes DynOPF-Net's inference time independent from the size of the power network, suggesting potential for large-scale systems}.

\section{Conclusion}\label{sec: conclude}
This paper was motivated by the need of developing a fast surrogate AC-OPF solver which takes into account the dynamic nature of a power network. While an increasing effort has been devoted to developing surrogate optimization models for AC-OPF, the paper shows their inability to deal with the system dynamics. We proposed DynOPF-Net, a novel integration of Learning to Optimize neural ODEs that successfully integrates synchronous generator dynamics within model training. 
To integrate the generator dynamics within the optimization, we employed a dual network architecture, consisting of a LtO model for estimating the OPF variables and neural ODE models to infer the dynamic behavior of each generator. Importantly, the proposed model is fully differentiable and allows for end-to-end training.
DynOPF-Net is compared with state-of-the-art LtO methods that have been proposed to solve the steady-state AC-OPF problem.
Empirical results report that all the baseline \LtO ~methods 
systematically fail to ensure system stability.
The findings pertaining to feasibility, optimality, and inference time consistently demonstrate that DynOPF-Net is capable of producing stable solutions, while achieving low steady-state gaps and minimal prediction errors, for real-time applications. 
Future work will focus on testing DynOPF-Net on larger power systems and will investigate the correlation between the prediction error of the OPF variables and violations of the stability constraints. 
Another line of work will focus on developing methods to ensure compliance with the system dynamics together with static AC-OPF constraint satisfaction.

\section*{Acknowledgments}

This research was partially supported by NSF awards EPCN-2242931, CAREER-2143706, and NSF awards 2041835, 2242930. 
Its view and conclusions are those of the authors only.

\bibliographystyle{unsrt}  
\bibliography{references}

\end{document}